# Multi-Objective Allocation of COVID-19 Testing Centers: Improving Coverage and Equity in Access

Zhen Zhong[1], Ribhu Sengupta[1], Kamran Paynabar [1], Lance A. Waller[2]

**Abstract**

At the time of this article, COVID-19 has been transmitted to more than 42 million people and resulted in more than 673,000 deaths across the United States. Throughout this pandemic, public health authorities have monitored the results of diagnostic testing to identify hotspots of transmission. Such information can help reduce or block transmission paths of COVID-19 and help infected patients receive early treatment. However, most current schemes of test site allocation have been based on experience or convenience, often resulting in low efficiency and non-optimal allocation. In addition, the historical sociodemographic patterns of populations within cities can result in measurable inequities in access to testing between various racial and income groups. To address these pressing issues, we propose a novel test site allocation scheme to (a) maximize population coverage, (b) minimize prediction uncertainties associated with projections of outbreak trajectories, and (c) reduce inequities in access. We illustrate our approach with case studies comparing our allocation scheme with recorded allocation of testing sites in Georgia, revealing increases in both population coverage and improvements in equity of access over current practice.

[1] H Milton Stewart School of Industrial and Systems Engineering, Georgia Institute of Technology, Atlanta, Georgia, USA
[2] Department of Biostatistics and Bioinformatics, Rollins School of Public Health, Emory University, Atlanta, Georgia, USA

# I. Introduction

In public health and epidemiology, diagnostic tests to identify diseased individuals and hotspots of transmission provide critical information for monitoring and understanding disease outbreaks within communities. Such information is essential for public health officials to estimate local prevalence and transmission, and to effectively plan for required treatment resources such as ICU beds, ventilators, personal protective equipment, and medical staff. Additionally, accurate estimates of the number of infected people can be used to develop probabilistic and statistical models to estimate the current effective reproduction numbers of the disease (measuring anticipated epidemic growth), and to predict likely spatial and temporal trajectories of the outbreak. As we have seen in the current COVID-19 pandemic, such data provide vital information for planning actions and policies, including recommended guidelines for social distancing, school openings and closures, remote work, community lockdowns, etc.

Despite the importance of testing and identification of positive cases, designing and implementing strategies for broad-scale testing have been and remain challenging tasks, particularly with respect to the overall management of testing logistics. At the beginning of the pandemic, limited availability of testing kits resulted in a struggle in allocation of and access to testing. Shifts in active variants of the virus within different regional populations also complicates matters generating sudden increases in testing needs as variants appear and spread in new regions. Finally, the decentralized nature of testing allocation involves both public agencies and private providers with little cross-communication, further muddying the waters. Such limitations signify a pressing need for a systematic systems approach for defining testing strategies to determine the number and location of satellite and mobile testing centers over time to best serve the at-risk populations and maximize the information obtained from test results. Strategic allocation of testing can improve accuracy of estimates of the number of cases at the local, state, regional, and national levels. Additionally, such strategies enable effective modeling and prediction of outbreak trajectories both in time and across geographical locations, thereby providing essential information for situational awareness and targeted public health response.

To propose an effective test site allocation scheme, we consider three factors: (i) how to maximize the numbers of people covered by allocated test centers (ii) how to facilitate the modeling and prediction of outbreak trajectories, and (iii) how to ensure equity in access to diagnostic testing between different sociodemographic groups. Maximizing nearby population coverage can reduce total travel distance to test centers of the entire population, thus increasing an individual's likelihood to seek a test. Facilitating modeling and prediction of outbreak trajectories allows targeting of prompt actions to block or reduce known transmission paths and adjust vaccine allocation schemes to prevent severe local outbreaks of COVID; while reducing inequity in access to test sites can reduce social injustices and health disparities associated with the disease.

Much research has been conducted addressing site allocation problems in public health [1-3,5-12]. Regarding population coverage, Cooper [1] addresses the site allocation problem by minimizing the total travel distance to the service location across all people. Narula and Ugonnaya [2] expand the framework to introduce a hierarchical structure for solving the site allocation problem. However, neither approach considers our second and third criteria (prediction of outbreak trajectories and minimization of access equity). In related work, Nobels et al. [3] introduces a criterion for measuring equity in spatial accessibility of pediatric primary healthcare access between children insured by Medicare and other children. However, since their application is not related to infectious disease, they do not specifically consider prediction of outbreak trajectories in their approach. To the best of our knowledge, other approaches [5-12] only consider one or two of our three criteria, limiting their ability to achieve the goals discussed earlier.

We build on these partial solutions to formulate the allocation problem as a multi-objective optimization model integrating all three criteria; namely, an approach to provide maximal covering of the population, minimal prediction error through the D-optimality criterion for Gaussian Process (GP) models, and minimal gap in test access among different demographic groups. We combine the three objectives using a set of weights determined by the user to indicate the relative importance of each objective. To solve the overall multi-criteria optimization problem, we propose a heuristic method using a genetic algorithm (GA) formulation.

The remainder of this paper is organized as follows. Section 2 provides an illustration of our methodology. Section 3 provides the details of our approach. Section 4 provides a case study showing our model's performance compared to current ground truths, and reveals important considerations regarding data availability and interpretation of results. Section 5 offers conclusions and directions for future research.

**II. Problem Definition**

As discussed in the introduction section, we will consider three key (possibly contradictory) criteria: (1) Maximizing a measure of population coverage denoted by $f_1$, (2) Minimizing the variance of prevalence predictions using a GP via D-optimality denoted by $f_2$, and (3) Minimizing sociodemographic gaps in test access denoted by $f_3$. We define the constraints based on the number, location, and capacity of available operational test sites (both static and mobile). We use a consolidated objective function based on a linear combination of $f_i; i = 1,2,3$ to create the optimization problem as follows

$$\max\{\lambda_1 f_1 - \lambda_2 f_2 - \lambda_3 f_3\}$$

$$subject\ to\ \text{limits on:}$$

the total number of operational test sites, and the capacity of each test site is limited.

In the following section, we will discuss the detailed formulation of $f_1$, $f_2$ and $f_3$ as well as the limits on the number and capacity of test sites.

### III. Mathematical Formulation

In section 3.1, 3.2, and 3.3, we will introduce the coverage criteria, the D-optimality criteria, and the equity criteria. We discuss the mathematical formulation of constraints in section 3.4, the final problem formulation in section 3.5, and the optimization approach in section 3.6.

### 3.1 Coverage criterion

Suppose there are *n* potential locations for test centers (denoted $d_i, i \in \{1, ..., n\}$) that should be used to provide maximum coverage to *m* areas (such as census blocks, census tracts or ZIP code areas) denoted by $c_j$, $j \in \{1, ..., m\}$. Define $e_j = 1$ if area *j* is covered by at least one test center, and $e_j = 0$, otherwise. We consider the area to be "covered" if a certain percentage of the area's population (set by the user) is covered by a test center. Then, the total coverage is defined as $f_1 = \Sigma_{j=1}^{n} e_j$.

## 3.2 D-optimality criteria.

The prevalence data obtained from test centers (number/proportion of positive tests) can be used to model the trajectory of the outbreak. We propose a general regression modeling and smoothing approach involving a Gaussian process (GP). Assume the spatial spread of the virus can be described by the following GP model, e.g.

$$y(x_1^*, x_2^*) = GP(x_1^*, x_2^*, \boldsymbol{\theta}) + \epsilon, \; \epsilon \sim N(0, \tau^2)$$

where $(x_1^*, x_2^*)$ represents the coordinate of locations, $y(x_1^*, x_2^*)$ represent the total number of COVID-19 cases detected in the area centered at $(x_1^*, x_2^*)$. The notation $GP$ represents a Gaussian process where $\boldsymbol{\theta}$ stands for parameters, e.g., mean, etc., estimated from the data. The precision of the GP prediction model depends on the uncertainty of the estimates of $\boldsymbol{\theta}$. We seek an allocation of test sites minimizing the uncertainty of $\boldsymbol{\theta}$ or, in other words, we seek to minimize the determinant of covariance matrix of $\boldsymbol{\theta}$.

It is known that for a Gaussian random field $F(\theta)^{\frac{1}{2}}(\hat{\theta}_n - \theta) \to^{Asymptotic} N(\boldsymbol{0}, \boldsymbol{I})$ where, $F(\theta)$ represent the Fisher information matrix of $\theta$ [4]. This relationship suggests we use $F(\theta)^{-1}$ to approximate the $cov(\theta)$, where

$$F_{j,k}(\theta) = \frac{1}{2} tr\{\Sigma(\theta)^{-1} \Sigma_j(\theta) \Sigma(\theta)^{-1} \Sigma_k(\theta)\},$$

and $\Sigma_j(\theta) = \frac{\partial \Sigma(\theta)}{\partial \theta_j}$, the derivative of the covariance matrix of $\theta$. The inverse Fisher information matrix is straightforward to compute and will asymptotically converge to the real covariance matrix. Therefore, instead of minimizing the determinant of the covariance matrix of $\theta$, we instead minimize the determinant of inverse Fisher information matrix, an approach known as meeting the local optimal design criterion [4]. Suppose due to limited resources, only $k$ facilities can be operational and $z_i$ is a binary variable indicating facility $d_i$ is operational or not. We define $V_0(Z, \theta) = -\log det F(Z, \theta)$, which is an evaluation of the uncertainty of $\theta$ under an allocation scheme indicated by $Z = \{z_i; i = 1, .., n\}$. As a result, when seeking to minimize prediction error, the best test site allocation scheme should result in the lowest $V_0$. Therefore, the local optimal design criterion should be:

$$S_M(\theta_0) = argmin_Z V_0(Z; \theta_0)$$

We note this criterion requires the availability of a preliminary estimate of $\theta$, denoted by $\theta_0$. In the beginning of the pandemic, such an estimate was not available due to limited prevalence data. Thus, we consider minimizing the worst-case scenario through a minimax optimal design criterion [4] defined by

$$S_M = argmin_Z \max_{\theta \in \Theta} V_0(Z, \theta)$$

Since, for different $\theta$ values, we will have different $argmin_Z V_0(Z, \theta)$, we adopt a criterion that removes the influence of different value of $argmin_Z V_0(Z, \theta)$. Specifically, we define

$$f_2 = V_1(z, \theta) = argmin_Z \max_{\theta \in \Theta}(V_0(Z, \theta) - V_0(S(\theta), \theta)$$

where $S(\theta)$ represent the local optimal design over $\theta$. The function $f_2$ is used as the D-optimal criterion in our multi-objective optimization.

### 3.3 Equity criterion

In this section, we introduce the equity measure used in the optimization model. Following [9], we claim equity is achieved when the probability of being tested given social and demographic factors including race and sex is equal to the probability of being tested unconditional of any socioeconomic factor. Mathematically, if $Y$ is the binary variable indicating being tested or not and $X$ is the set of observed social and demographic factors, then our definition of equity can be measured by comparing the expectation of the marginal distribution of $Y$ to that of the conditional distribution $Y|X$. Since $Y$ is an indicator function, the expectation of marginal distribution $E(Y)$ is equal to $P(Y)$ and the expectation of conditional distribution $E(Y|X)$ is equal to $P(Y|X)$. To move toward an equitable allocation of testing resources, we minimize the quadratic loss between $P(Y|X)$ and $P(Y)$. Therefore, we define the following equity criteria as:

$$\Sigma_{v_1=1}^{V_1} \ldots \Sigma_{v_u=1}^{V_u} \left( \Sigma_j \left( \frac{P_{v_1\ldots v_u,j} e_j}{\Sigma_j P_{v_1\ldots v_u,j}} \right) - \Sigma_j \left( \frac{P_j e_j}{\Sigma_j P_j} \right) \right)^2$$

Here $v_1, \ldots v_u$ represent labels for sociodemographic subpopulations, while $V_1, \ldots V_u$ represent the total number of different labels in each group, $\left( \frac{P_{v_1\ldots v_u,j} e_j}{\Sigma_j P_{v_1\ldots v_u,j}} \right)$ is the coverage probability for a specific sociodemographic group, while $\left( \frac{P_j e_j}{\Sigma_j P_j} \right)$ yields the probability to be covered among all different sociodemographic groups. Ideally, these two quantities should be the same, and we define or criteria for optimizing equity in test site allocation as the mean square loss of the difference between these two probabilities.

### 3.4 Model Constraints

As discussed in the problem formulation, we consider two types of constraints. The first constraint, defined by $\Sigma_i z_i \leq k$, indicates the number of operational test sites is at most $k$ (where $k$ is determined based on available recourses, i.e., how many test sites can we afford?). The second constraint pertains to the capacity of each test site and its impact on providing access to individuals living in the surrounding areas. Define the binary variable $a_{ij}$, where $a_{ij} = 1$ if residences of area

$c_j$ can be covered by facility $d_i$, and $a_{ij} = 0$, otherwise. Then, we define the capacity constraint by $\Sigma_i a_{ij} z_i - e_j \geq 0$. This constraint implies that if there is not enough capacity in facility $d_i$ to cover area $c_j$ (i.e., $a_{ij} = 1$), or this facility is not selected, then the area $c_j$ is not covered (i.e., $e_j = 0$).

In practice, we need to determine the matrix $A = \{a_{ij}\}$. This matrix reflects the location and coverage (by the site-specific testing capacity, number of tests available each day) of each test site and can be obtained by using following algorithm.

**Table 1. Algorithm for Determining $A$ matrix**

| | |
|---|---|
| Step 1 | Based on the testing capacity of each test site, we define the total population that each test site can cover, denoted as $TP$. Also, we denote the population of area $j$ by $P_j$. |
| | **While** (Not all candidate test sites are analyzed): |
| Step 2 | The location of potential test site $i$ is denoted by $(x_{i1}, x_{i2})$, which is the centroid of an area. We calculate the distance between test site $i$ and area $j$, and define it as $D_{ij}$. |
| Step 3 | Rank the $D_{ij}$ from ascendingly, with the corresponding subscript $i_{(j,1)}, \dots, i_{(j,m)}$. |
| Step 4 | Find $r_i$ such that $\Sigma_{l=1}^{r_i} w_{i_{(J,l)}} P_{i_{(J,l)}} \leq TP < \Sigma_{l=1}^{r_i+1} w_{i_{(J,l)}} P_{i_{(J,l)}}$. Mark $a_{i_{(J,1)}}, \dots, a_{i_{(J,l)}} = 1$, and any other $a_i$ is set to zero. $w_{i_{(J,l)}}$ is a weight for the population which is always set to 1. |
| | **End** |

For the above algorithm, we assume people always choose the closest test site to their home to be tested and we claim one area is covered if the population size of the area is below the capacity of the site.

To handle various test sites with different capacities or types of test that they provide, we generalize the constraint to $\sum_{t=1}^{t_0} \sum_{i=1}^{m} a_{ij}^{(t)} z_i - e_j \geq 0$, where $a_{ij}^{(t)}$ denotes the $(i,j)^{th}$ element of the $A$ matrix of type $t$ test sites and $t_0$ represents the total number of test site types. This holds since a person is covered if covered by at least one type of test site.

### 3.5 Final problem formulation.

Taken together the objective and constraints, our problem formulation becomes:

$$\max\{\lambda_1 f_1 - \lambda_2 f_2 - \lambda_3 f_3\}$$

$$= \max\left\{\lambda_1 \sum_{j=1}^{m} e_j - \lambda_2 V_1(z,\theta) - \lambda_3 \Sigma_{v_1=1}^{V_1} \ldots \Sigma_{v_u=1}^{V_u} \left(\left(\frac{P_{v_1\ldots v_u,j} e_j}{\Sigma_j P_{v_1\ldots v_u,j}}\right) - \left(\frac{P_j e_j}{\Sigma_j P_j}\right)\right)^2\right\}$$

$s.t.$

$$\sum_{t=1}^{t_0} \sum_{i=1}^{m} a_{ij}^{(t)} z_i - e_j \geq 0 \; ; \; \sum_{i=1}^{m} z_i = k; e_j \in \{0,1\}; z_i \in \{0,1\}.$$

### 3.6 Optimization approach

The proposed optimization model is a non-convex optimization problem constrained to integer solutions. As the size of the problem can become very large for some states, applying exact integer programming algorithms may prove computationally infeasible, and we explore the use of a genetic algorithm approximation approach to solve the optimization problem. To solve this model, we use the Genetic Algorithm function from the gramEvol package in R. We define the number of chromosomes as the total number of test centers to be allocated and the best genome result to be the solution. The fitness function is our objective function.

**IV. Case study**

To illustrate and validate the proposed allocation approach, we apply it to data from multiple counties in the state of Georgia (as of February 22, 2021) and compare our method with current testing allocation schemes observed. For simplicity, we treat each census tract as one area and define the test site capacity as 1,120 tests per week. We choose to measure our coverage at the census tract level because census tracts typically have similar total populations. Our data on existing test center locations is a crowd-sourced dataset from the URISA GISCorps. Volunteers from health departments, local governments, and healthcare providers have been updating this website daily with existing test site information across the country with exact geo-locations for each test site, information about hours, public vs private funding/ownership, and many other details [13]. For our case study, given that our model is built for public officials to allocate public test sites, we filter all existing test sites for only public test sites. For census tract information, we are using the 2010 Census results for the state of Georgia. As of the 2010 Census, the state of Georgia has a population of 9,687,653 people with 1,969 census tracts.

We begin with the available allocation of testing centers on February 22, 2021 within each county in Georgia, based on our data and compare this allocation with the proposed allocation of the same number of testing sites allocated according to our three criteria: (1) the coverage score: $\sum_{j=1}^{m} e_j$,

(2) the D-optimality score $V_1(S,\theta)$, and (3) the equity score $\Sigma_{v_1=1}^{V_1} \ldots \Sigma_{v_u=1}^{V_u} \left(\left(\frac{P_{v_1\ldots v_u,j}e_j}{\Sigma_j P_{v_1\ldots v_u,j}}\right) - \left(\frac{P_j e_j}{\Sigma_j P_j}\right)\right)^2$. Note that larger coverage scores are preferred, while for the D-optimality and equity scores, the lower the better. Setting our weights as $\lambda_1 = 10^{-2}$ and $\lambda_3 = 1$, we obtain the results shown in Table 1. For our target population percentage, we set it to 10% meaning a census tract will be covered if the testing center can cover 10% of its population. Further discussion about the choice of these weights is discussed below.

Table 1. Case Study Results

|  | Coverage Score | D-Optimality Score | Equity Score |
|---|---|---|---|
| **Current scheme (Fulton)** | 33 | $8.97 \times 10^{-5}$ | 0.162 |
| **Proposed scheme (Fulton)** | 46 | $2.35 \times 10^{-5}$ | 0.153 |
| **Current scheme (Cobb)** | 16 | $1.2 \times 10^{-4}$ | 0.179 |
| **Proposed scheme (Cobb)** | 25 | $1.153 \times 10^{-4}$ | 0.175 |
| **Current scheme (Gwinnett)** | 5 | $2.07 \times 10^{-4}$ | 0.208 |
| **Proposed scheme (Gwinnett)** | 5 | $4.763 \times 10^{-4}$ | 0.189 |
| **Current scheme (De Kalb)** | 32 | $2.07 \times 10^{-5}$ | 0.172 |
| **Proposed scheme (De Kalb)** | 50 | $3.722 \times 10^{-5}$ | 0.169 |
| **Current scheme (Chatham)** | 18 | $4.4 \times 10^{-4}$ | 0.250 |
| **Proposed scheme (Chatham)** | 37 | $3.525 \times 10^{-4}$ | 0.158 |

From the above table, we find our algorithm always generates better coverage and equity scores than the observed distribution of test sites for all counties considered. However, for the D-optimality score, our approach does not always perform as well as the current testing distribution, suggesting competition between our optimality criteria. That is, simply improving coverage can make equity worse. Our solution considers the weights assigned to the equity and coverage criteria. In the case study, both weights are set to $10^{-2}$. However, if we wish to improve the D-optimality criteria, we can simply increase or decrease the values of our weights $\lambda_1$ and $\lambda_3$.

Overall, the higher the value given to each weight, the more importance it takes in the final solution. For choosing weights, our initial analyses for the state of Georgia suggest that the coverage score's magnitude ranges from $10^0$ to $10^1$, the D-optimality score's magnitude ranges from $10^{-5}$ to $10^{-4}$, and the equity score's magnitude is typically $10^{-1}$. With these magnitudes in mind, we suggest the user changes the weight of $\lambda_1$ to increase/decrease the weightage of coverage. For decision makers, we assign four general levels of importance (in the order of Less Important, Somewhat Important, Important, and Very Important) for both weights, and suggest the following illustrative values for each in Table 2.

Table 2. Importance Levels Recommended for Weights

| Weight | Less Important | Somewhat Important | Important | Very Important |
|---|---|---|---|---|
| $\lambda_1$ (Coverage) | $10^{-6}$ | $10^{-4}$ | $10^{-2}$ | $10^0$ |
| $\lambda_3$ (Equity) | $10^{-4}$ | $10^{-2}$ | $10^0$ | $10^2$ |

We do not recommend setting either weight outside of the magnitudes of $[10^{-8}, 10^6]$ for consistent model performance. Please note these weights and recommendations are for use cases in the state of Georgia, these values should be updated if the state is changed. Future work will define more generalized weights to allow broader application to other settings. A web-based app has been developed using which users can run our model for different counties. Users are also able to change the weights and compare the results with respect to the three criteria [14[3]].

**4.1 Case Study Results Insights**

Figure 1 illustrates a comparison of the February 22, 2021 allocation of reported public test sites and the allocation of public testing for Fulton County defined by our algorithm. The Fulton County results provide deeper insights into the features driving the current allocation of testing sites. The results illustrate that, for a better equity score, more test centers should be moved from the urban centers of Atlanta in Fulton County to the suburbs in the greater Atlanta region in areas such as Alpharetta and Johns Creek in the norther part of the county. Mathematically, the model result creates a better equity score as the urban centers of Fulton County are predominantly non-White and these suburbs are predominantly white, a setting we consider in more detail below. For a general map and a map of the racial breakdown for Fulton County at the census tract level, please reference Appendix A.

---

[3] [COVID-19 Test Center Allocation Tool](#)

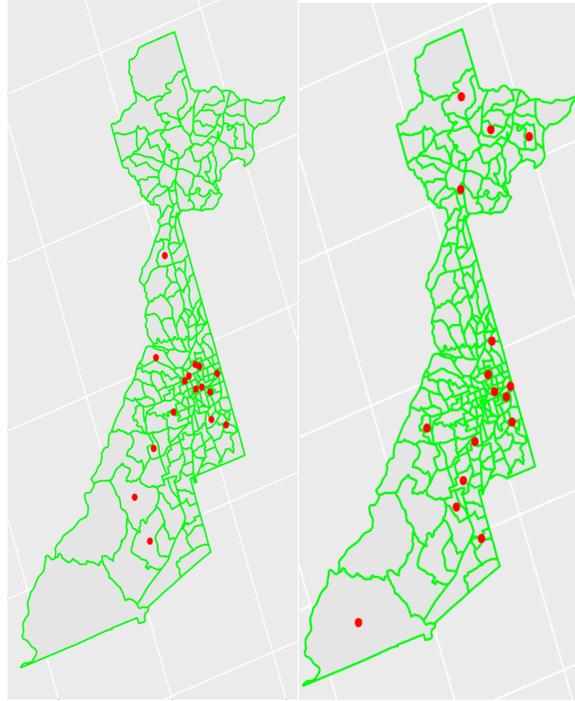

**Figure 1. Current (Left) and Suggested (Right) Allocation of Test Centers for Fulton County**

These results merit a closer look. As noted above, our model is focused on *public* testing sites, which means common *private* testing sites such as CVS/Walgreens and doctors' offices are not considered. Keeping in mind the well-established inequities in regular health care access between non-white urban areas and predominantly white suburbs, our data (and hence, our model) does not consider preexisting availability of private testing in suburban areas [15]. Our crowdsourced data (which is incomplete) show us that in these areas where our model is proposing to move some centers to, there are already many private options available. A map of an incomplete list of private test centers in Fulton County can be seen in Figure 2 illustrating many preexisting private test centers in the areas the model suggests moving test centers to for equity purposes, when only based on existing public testing sites.

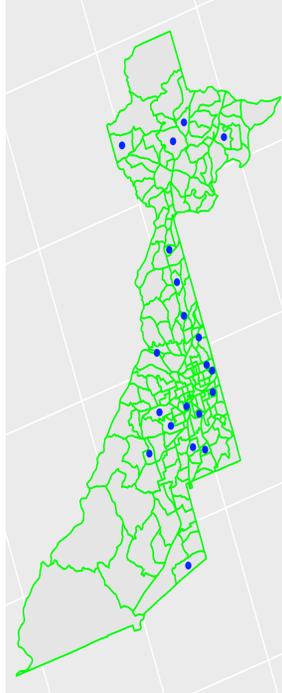

**Figure 2. Current Allocation of Private Test Centers in Fulton County**

The Fulton County example illustrates a need for comprehensive data regarding all available sources of testing for the proposed optimization to have greatest impact, and a potential for the optimization results to compound existing inequities, if based on incomplete data.

We conduct a similar case study on Cobb County. A comparison of the current allocation and the suggested allocation of public testing sights for Cobb County can be seen below in Figure 3. Compared to Fulton County, Cobb County is comprised of more suburban areas and not many urban centers. The model decides to spread out testing sites across the county and is choosing census tracts which have close to 50% White vs 50% non-White racial distribution. This allocation lowers the equity score compared to the current allocation. The model is also choosing some census tracts strategically to help increase the number of census tracts covered due to the high weight given to coverage in our case study.

Similar to Fulton County, the incomplete crowd-sourced data available on existing test sites shows that there is a large number of private test sites allocated in Cobb County. The allocation of these private test sites is visible in Figure 4. The high number of private test sites compared to public test sites is presumably due to the high number of suburban areas such as Smyrna, Marietta, and Kennesaw. In fact, most of the private test centers on the map fall within and around these suburbs. This large difference also highlights the difference in approaches a county could take in terms of reliance on public or private testing sites to fulfill testing demands.

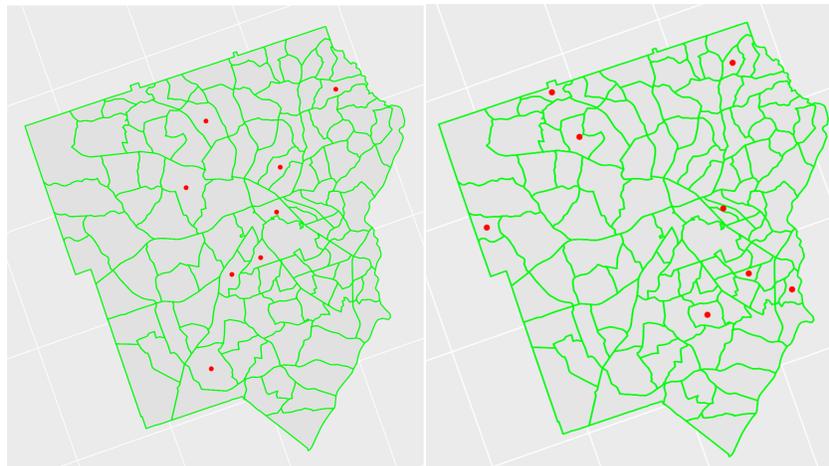

**Figure 3. Current (Left) and Suggested (Right) Allocation of Test Centers for Cobb County**

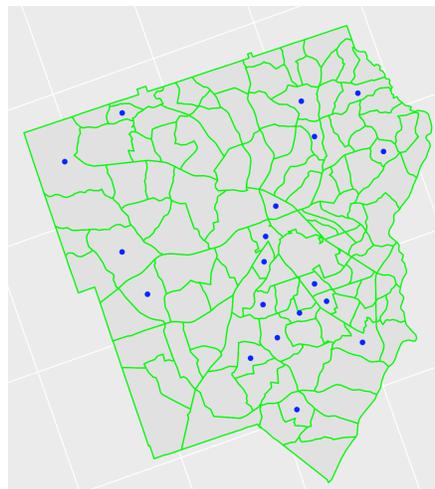

**Figure 4. Current Allocation of Private Test Centers in Cobb County**

To expand our scope beyond the Greater Atlanta Area, we also consider Chatham County, home to the city of Savannah, with results visible in Figure 5. The Chatham County results show that the current allocation of public testing sites are clustered in the main urban area of Savannah and there are no public centers allocated to the suburbs of Savannah. Based on the large allocation of private centers in Fulton and Cobb County for Atlanta suburbs, upon checking the allocation of private centers in Chatham County, we found that there were only three total private test centers in Chatham County according to the crowdsourced data. The placement of these private sites can be seen in Figure 6.

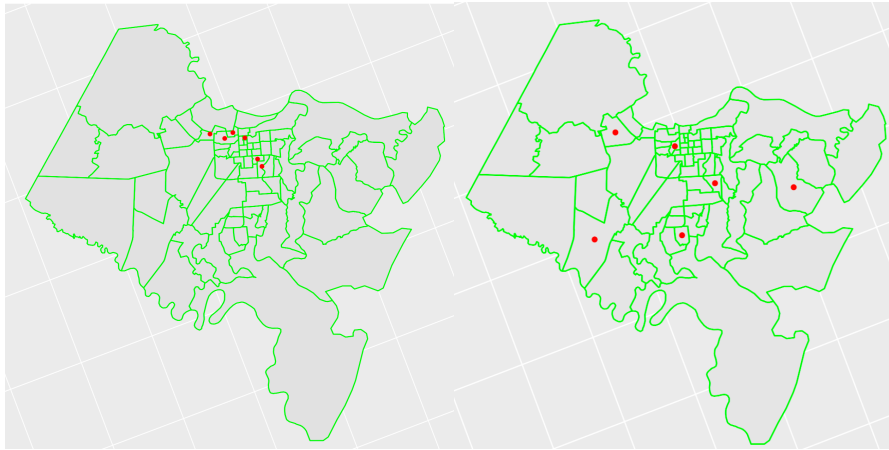

**Figure 5. Current (Left) and Suggested (Right) Allocation of Test Centers for Chatham County**

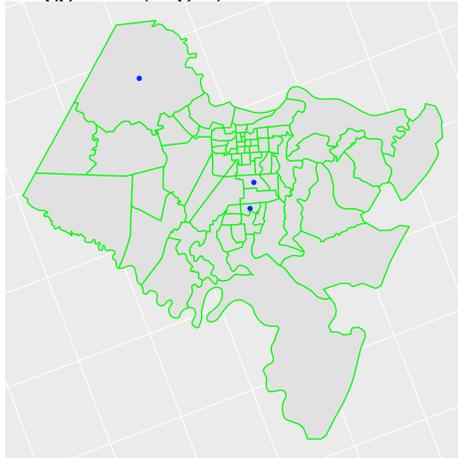

**Figure 6. Current Allocation of Private Test Centers in Chatham County**

Overall, there seems to be a small number of test sites in Chatham County which raises important limitations in using our model. When there are a small number of test sites to be allocated, the equity portion of the optimization model will try and place test sites into areas which are more evenly split along socioeconomic strata..

Similar to the Fulton County results, the model decides to moves test centers from predominatly non-white areas and put them in areas with more sociodemographically balanced populations (for a map with racial breakdowns of Chatham County, please reference Appendix C). The proposed allocation provides important insight in the application of a criteria seeking to minimize an equity score: If the allocation of public sites already favors non-white populations, perhaps to fill in gaps in coverage by public test sites, minimizing our equity score *based only on public sites* will move sites to more a racially balanced allocation. However, if the white population has access to additional private testing opportunities not included in the data, our optimization model can (and, often, does) compound existing inequities by proposing adjustments of public site locations away from non-white populations. Future work will extend the proposed approach to include both public and private test sites, but will require more comprehensive data on test sites to avoid the biases seen in public-only results.

The public-only allocations illustrated above seem to be hindering access for historically underserved communities of color, especially in urban centers in cities like Atlanta and Savannah. However, the combination of reliance on private testing centers in suburban areas (often predominatnly White) and the overall low number of test centers available require public-only model results to be interpreted with caution, especially when used with incomplete data. For example, in Chatham County, given the low total number of public or private test sites listed in the crowdsourced data, our model results should give public officials an indication that there is a need for more test centers. As more test centers are added, the model would place more test centers in areas where the testing need is high in urban areas, and would do so in a manner to still minimize the inequity in access.

## 4.2 Case Study Discussion

These results above raise an important point for such multicriteria optimization models used for test center allocation. This allocation happens within a context of the current modus operandi for allocating public resources. Some states may pursue a public-first philosophy where the main objective is to give all residents easy access to public testing sites. From there the state can forge partnerships with private firms to fill in the gaps in availability. On the other hand, states could use public sites as a second resort after all private testing centers are established. In this case, the public allocation would be used to fill in the gaps in availability provided by private sites, often resulting in allocation of public sites to urban areas with higher proportions of non-white residents. Based on the data in our case study, this seems to be the case for many counties in Georgia. Such allocations compensate for inequities in location of private testing facilities, but, as we see above, by applying our optimization procedure to only the public test sites and requiring equity in access, the algorithm undoes the preferential assignment and proposes reallocation of public sites *away* from the very neighborhoods where they were placed to help. Future considerations will consider extensions to better include original goals as well as criteria for future allocation of both private and public test sites.

Expanding this discussion, decision makers must also consider many socioeconomic and geographic factors when allocating resources. For example, suburban residents usually own a motor vehicle and can travel longer distances to get access to testing and utilize larger testing sites with capabilities such as drive-through testing. In contrast, urban residents may be more reliant on public transportation and thus must have testing centers which are within a walkable distance or are located in higher density neighborhoods that do not have the space for larger drive-through testing centers. This context is critical for understanding the examples above and for further refining the approach. Eventually, extra constraints can be added to this model to allow for certain areas to not receive testing sites (as they could already be serviced by private sites), to allow for differences in transportation access, and to allow for multiple test-center types to be allocated. Note that currently the web-based app allows for two test center types, regular test centers and mega test centers.

Another key aspect of our model is data support, i.e., the availability and type of information regarding possible test sites going into the model. With the issue of the mixture of private and public testing sites, the user of the model must know as much of this context as possible to get an allocation which is practical. If the user does not know the current allocation of public and private testing sites, they could end up with an allocation where public and private testing sites are placed side by side. With COVID-19 testing declared to be free and having no insurance requirements through the Families First Coronavirus Response Act, this would be a waste of public facilities and funds [14]. Overall, the lack of a centralized data source of all available testing locations is a big hinderance to optimizing access to all. The Cobb County results illustrate this point well. It is evident there is not much of a need for public testing locations in Cobb County, however for a public official to decide where to put public site, it is imperative they already know where the current private locations are. Also, our Chatham County results show how allocation with limited number of test sites can lead to conundrums for public officials with allocations which do not give enough access to underserved communities. While there could very well be a low number of test sites available in the city of Savannah, it seems very unlikely that only three private test centers would be open during a peak period of the epidemic. This type of data access problem can only be addressed with a consolidated data source available to our model user including the current allocation of both public and private testing centers. Our model would be best used in a scenario where all data on all active test sites, public and private, are available readily for users.

## V. Conclusion

This paper proposes a novel test site allocation scheme by formulating a multi-objective optimization problem. The objectives are (i) increase in coverage; (ii) reduction of prediction uncertainties; and (iii) improvement of equity among different social demographic groups. With this framework we have built a model and an interactive tool which can take these objectives into consideration and deliver a Covid-19 Test Center allocation for a given number of test centers and their capacities to public health officials. Our tool allows the user to choose the percentage of the population they want to cover and the relative importance of each component in the optimization problem. Most importantly, our methodology allows for optimal allocations within the context of minimizing inequity in a given allocation. Using our model compared to existing location allocation algorithms in practice, our allocation scheme can outperform current test site allocation schemes. However, for our models to be used accurately, consolidated data sources containing full knowledge of existing public and private test centers are crucial for achieving best results. As the examples illustrate, ignoring existing differences in access and placement of public and private test centers can result in increasing overall inequity while attempting to minimize inequities for public sites alone. Moving forward, our model can serve as a baseline for many public health decisioning frameworks in the future to not only deliver optimal allocations, but allocations which help tackle historic inequities within society.

# Appendix A

Fulton County Map

(https://www.mapsofworld.com/usa/states/georgia/counties/fulton-county-map.html)

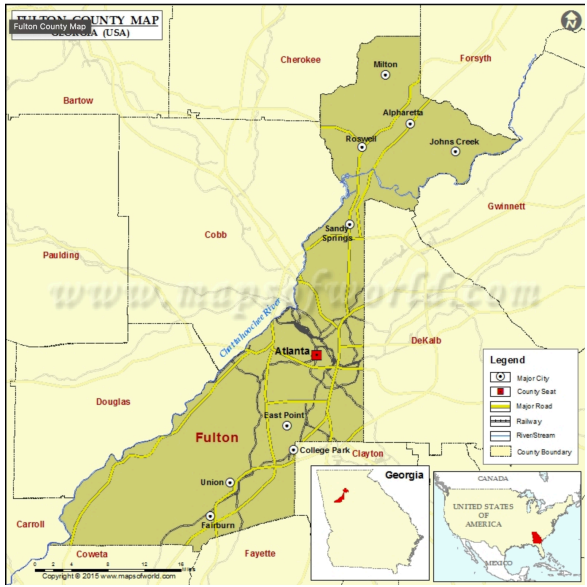

Fulton County Racial Breakdown

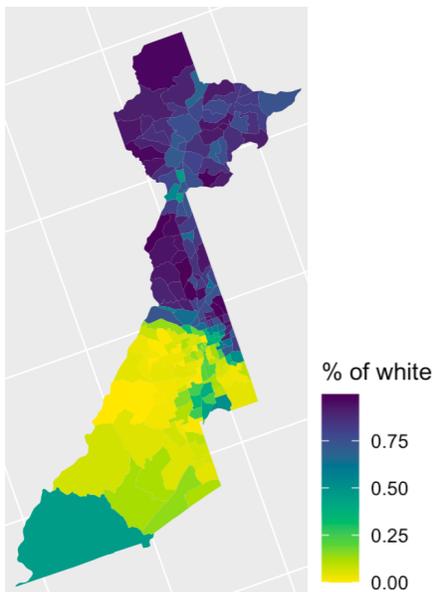

# Appendix B

Cobb County Map

https://www.mapsofworld.com/usa/states/georgia/counties/cobb-county-map.html

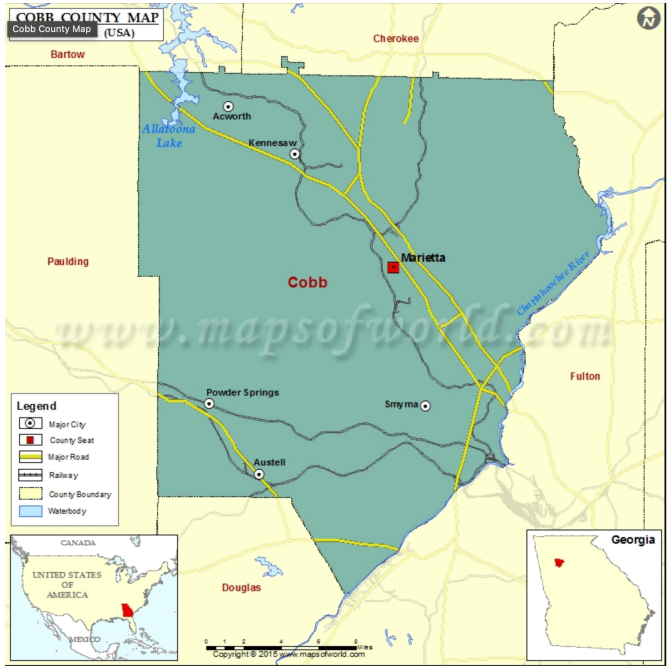

Cobb County Racial Demographics

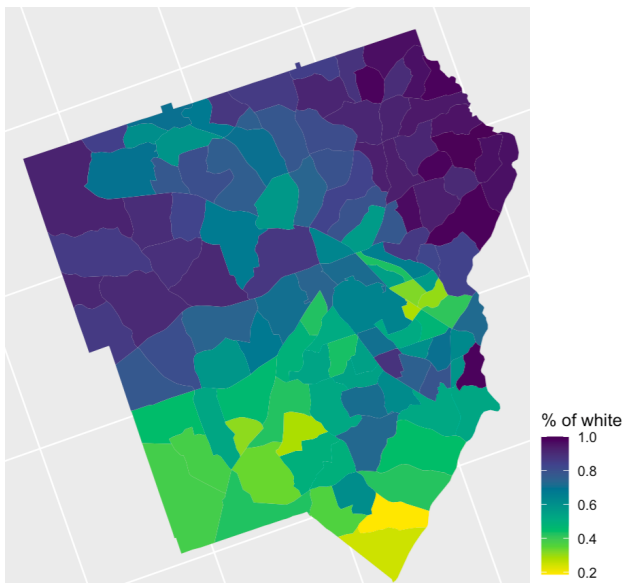

# Appendix C

Chatham County Map

https://www.mapsofworld.com/usa/states/georgia/counties/chatham-county-map.html

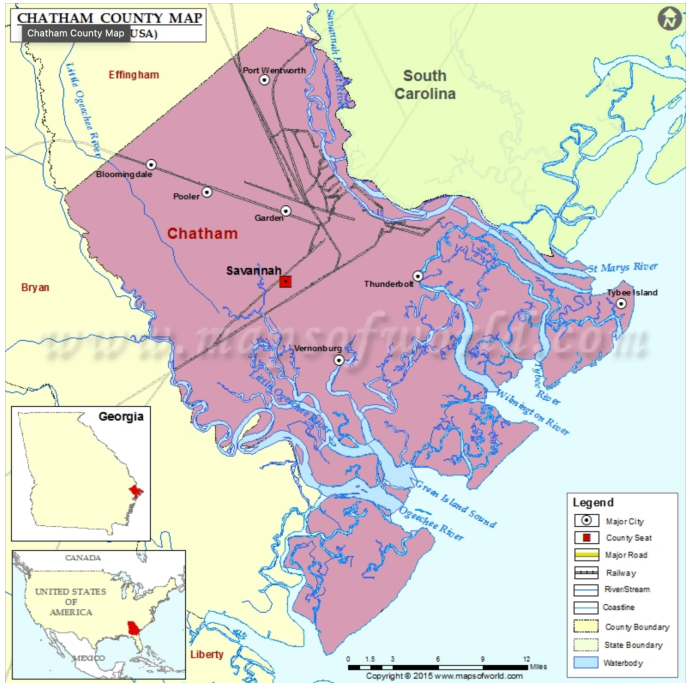

Chatham County Racial Demographics

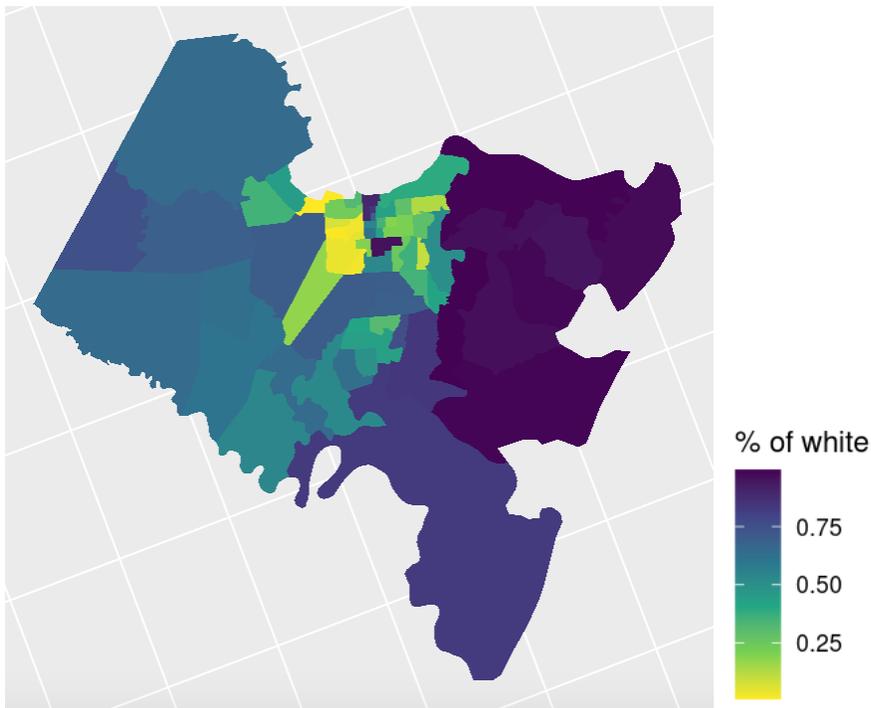